# An iterative nonlocal residual constitutive model for nonlocal elasticity

**Mohamed Shaat∗**

*Engineering and Manufacturing Technologies Department, DACC, New Mexico State University, Las Cruces, NM 88003, USA*

*Mechanical Engineering Department, Zagazig University, Zagazig 44511, Egypt*

**Abstract**

Recently, it was claimed that the two-phase local/nonlocal constitutive models give well-posed nonlocal field problems and eliminates the ill-posedness of the fully nonlocal constitutive models. In this study, it is demonstrated that, both, the fully nonlocal and the two-phase local/nonlocal constitutive models secrete ill-posed nonlocal boundary value problems. Moreover, it is revealed that all Eringen's integral and differential nonlocal constitutive models secrete unsolvable nonlocal boundary value problems.

In this study, it is demonstrated that solutions of nonlocal elasticity problems are exist, and Eringen's constitutive model cannot determine these solutions. To overcome the limitations of Eringen's constitutive models, novel integral and differential iterative nonlocal residual constitutive models are proposed. Using these two constitutive models, the sum of the nonlocal residual field at a point is iteratively formed. Then, this nonlocal residual is imposed to the local boundary value problem. Thus, the nonlocal elasticity is obtained in the form of a local boundary value problem with an imposed nonlocal residual field. Using any of these constitutive models, a solution is guaranteed for a nonlocal field problem. To show the effectiveness of the proposed constitutive models, the nonlocal field problems of beams with different natural boundary conditions are considered. The results of the proposed integral and differential constitutive models are identical and feasible.

## 1. Introduction

Eringen proposed the nonlocal continuum theory that can model long-range interatomic interactions between material particles forming the continuum [1-7]. In the context of the nonlocal theory, the balance equations are with the form the same as the balance equations of the classical theory. However, the constitutive equations of the nonlocal theory depend on operators that sum the neighbor (local) and non-neighbor (nonlocal) interactions. Thus, in the context of the nonlocal theory, the nonlocal field is considered as the sum of the local field and a nonlocal residual [1-3]. To incorporate these nonlocal residuals in the framework of linear elasticity, a constitutive model that depends on an integral operator was proposed [4-7]. Then, complications were revealed when solving nonlocal field problems using this integral nonlocal constitutive model. Therefore, Eringen proposed a differential nonlocal constitutive model that depends on a differential operator [8].

---

∗Corresponding author. Tel.: +15756215929.
 *E-mail addresses:* shaat@nmsu.edu; shaatscience@yahoo.com (M. Shaat).

Recently, discrepancies were revealed between solutions of nonlocal field problems formed using Eringen's differential nonlocal constitutive model and integral nonlocal constitutive model [9-18]. For instance, solutions of some bending of beams problems modeled using Eringen's integral nonlocal constitutive model were obtained identical to the classical solutions [9-11]. In addition, it was observed that cantilever beams exhibit hardening behaviors when modeled using the differential nonlocal constitutive model and softening behaviors when modeled using the integral constitutive model [12]. The paradoxes in the existing results of the differential nonlocal constitutive model were discussed in [12-18]. The observed paradoxes and discrepancies between Eringen's two constitutive models were attributed to the inconsistency between the nonlocal natural boundary conditions of the nonlocal field problem and the boundary conditions of the constitutive model [12, 15, 19]. Thus, the equivalence between the integral constitutive model and the differential constitutive model is constrained to be hold only if both models satisfy the constitutive boundary conditions [15, 19, 20].

Afterwards, Romano et al. [19] demonstrated that nonlocal beam problems formed using both Eringen's differential and integral nonlocal constitutive models admit either a unique solution (*for exceptional cases*) or no solutions at all. They demonstrated that a nonlocal beam problem is ill-posed and unsolvable when using either the integral constitutive model or the differential constitutive model. To overcome the ill-posedness of nonlocal elastic problems, Romano et al. [19, 20] suggested the use of the local/nonlocal mixture constitutive model [2, 21]. *In this study, it is demonstrated that the local/nonlocal mixture constitutive model is ill-posed and may give unsolvable nonlocal field problems as well (see Model 3 in section 2).*

In section 2, Eringen's constitutive models are discussed and examined according to their ability to secrete nonlocal field problems that admit solutions. It is demonstrated that all Eringen's constitutive models likely give nonlocal field problems that admit no exact solutions. In exceptional case, however, Eringen's nonlocal constitutive models give the same unique solution. In section 3, it is demonstrated that a solution of a nonlocal field problem should exist, but Eringen's constitutive models fail to determine it.

The main contribution of this study is to develop an iterative nonlocal residual constitutive model that effectively overcomes the paradoxes and the ill-posedness of Eringen's nonlocal constitutive models. Using the iterative nonlocal residual constitutive model a solution of the nonlocal field problem is guaranteed. According to the demonstrations provided in section 2, it is concluded that any attempt to solve the nonlocal field problem for the nonlocal field is bound to fail. This is because, to solve the nonlocal boundary value problem, the nonlocal boundary conditions should be correctly formed and satisfy the constitutive boundary conditions, and the existing constitutive models fail to fulfill this essential condition. Therefore, in this study, the author proposes *correcting the solution of the local field problem for the nonlocal residual field of the elastic domain*. Thus, in the context of the proposed approach, *the local field problem of an elastic domain with a nonlocal-type stress residual is solved with the local-type boundary conditions are applied*. First, the sum of the residual energy at a specific point due to its nonlocal interactions with the surrounding points is iteratively computed. Thus, the nonlocal residual energy is formed (***not determined***) depending on a pre-determined local-type field. Then, this iteratively-formed nonlocal residual energy is imposed to the local field problem. Afterwards, the field equation is solved in the local field with the imposed nonlocal residual field.

The proposed iterative nonlocal residual constitutive model is developed in section 3. Two forms of the constitutive model are derived such that one form depends on the integral operator and its corresponding differential form is, then, derived. An algorithm of solution is proposed to obtained solutions of nonlocal field problems using the iterative nonlocal residual constitutive model. Case studies of Euler-Bernoulli beams with different natural and essential boundary conditions are considered using the proposed constitutive model. It is demonstrated that both the differential and integral forms of the iterative nonlocal residual constitutive model give the same results. Thus, it is demonstrated that the proposed constitutive model provides solutions of nonlocal field problems with neither paradoxes nor ill-posedness are secreted.

## 2. Eringen's nonlocal constitutive models may not secrete solutions

As previously mentioned, the paradoxes and ill-posedness of Eringen's differential and integral constitutive models are discussed in some studies [9-18]. Romano et al. [19, 20] suggested the local/nonlocal mixture constitutive model [2, 21] to overcome the ill-posedness of Eringen's fully nonlocal constitutive models. In this section, it is demonstrated that all Eringen's constitutive models (*including the local/nonlocal mixture constitutive model*) likely provide nonlocal field problems which are either unsolvable or admit no solutions.

Consider an elastic continuum with a volume, $V$, and a surface, $S$. Hamilton's principle can be written for this continuum body as follows:

$$\int_0^{t_0} \left( -\int_V \left( \rho \frac{\partial^2 u_i}{\partial t^2} \delta u_i \right) dV + \int_V t_{ji,j} \delta u_i \, dV - \int_S n_j t_{ji} \delta u_i \, dS + \int_V f_i \delta u_i \, dV \right.$$
$$\left. + \int_S \bar{P}_i \delta u_i \, dS \right) dt = 0 \tag{1}$$

where $u_i$ denotes the displacement field. $f_i$ is a body force vector, and $n_j$ denotes the unit normal vector. $\rho$ is the mass density. $t_{ji}$ is the stress tensor, and $\bar{P}_i$ denotes the surface tractions vector.

According to Eringen's nonlocal theory [1, 2], the stress field, $t_{ji}$, in equation (1) is a nonlocal stress which accounts for the neighbor and non-neighbor interactions between points forming the continuum [22, 23].

According to equation (1), the balance equation is defined as follows:

$$t_{ji,j} + f_i = \rho \frac{\partial^2 u_i}{\partial t^2} \tag{2}$$

with the natural boundary conditions:

$$n_j t_{ji} = \bar{P}_i \tag{3}$$

It should be mentioned that the natural boundary conditions are nonlocal-type boundary conditions which depend on the nonlocal stress tensor, $t_{ji}$.

Eringen proposed different nonlocal constitutive models to define the nonlocal stress, $t_{ji}$. In this section, Eringen's nonlocal constitutive models are reviewed, and their abilities to secrete nonlocal field problems with feasible solutions for the nonlocal fields are discussed.

**Model 1: Nonlocal integral constitutive model**

By assuming a rapid attenuation of the nonlocal field, Eringen proposed a single attenuation function for all the material coefficients to accumulate the non-neighbor interactions in an elastic domain. Thus, for linear elastic continua, Eringen proposed the following constitutive model of the nonlocal stress [8]:

$$t_{ij}(x) = \int_V \alpha(|x' - x|) \, \sigma_{ij}(x') \, dV(x') \tag{4}$$

where

$$\sigma_{ij} = \lambda \varepsilon_{rr} \delta_{ij} + 2\mu \varepsilon_{ij} \tag{5}$$

where $\sigma_{ij}$ is the local-type stress tensor. $\varepsilon_{ij} = \frac{1}{2}(u_{i,j} + u_{j,i})$ is the infinitesimal strain tensor. $\lambda$ and $\mu$ are Lame constants.

In equation (4), the nonlocal stress is represented as a convolution law that depends on the kernel $\alpha(|\mathbf{x}' - \mathbf{x}|)$. This kernel is defined to emulate the attenuation of the nonlocal interactions with the distance $|\mathbf{x}' - \mathbf{x}|$. Eringen [8] proposed the special kernel function of the Green's function type, $\alpha(|\mathbf{x}' - \mathbf{x}|)$, as follows:

$$\alpha(|\mathbf{x}' - \mathbf{x}|) = \frac{1}{2\ell} \exp\left(-\frac{|\mathbf{x}' - \mathbf{x}|}{\ell}\right) \tag{6}$$

where $\ell$ is a characteristic length that can be defined to satisfy the following conditions:

$$\begin{cases} \int_V \alpha(|\mathbf{x}' - \mathbf{x}|) \, dV(\mathbf{x}') = 1 \\ \ell \to 0 \Rightarrow \alpha(|\mathbf{x}' - \mathbf{x}|) = \delta(|\mathbf{x}' - \mathbf{x}|) \end{cases} \tag{7}$$

where $\delta(|\mathbf{x}' - \mathbf{x}|)$ is the Dirac delta function.

To solve the nonlocal field problem (equations (2) and (3)) using Eringen's nonlocal integral constitutive model (equation (4)), an algorithm can be proposed, as follows:

> **Algorithm 1:**
>
> **Algorithm 1.1: Nonlocal field problems with natural boundary conditions**
>
> *The balance equation (2) is solved for the nonlocal stress field, $t_{ij}$, with the nonlocal-type boundary conditions (equation (3)) are applied. Then, the obtained nonlocal stress field is substituted into the constitutive model (equation (4)) which, then, can be solved for the local-type stress field, $\sigma_{ij}$.*
>
> **Algorithm 1.2: Nonlocal field problems with essential boundary conditions**
>
> *The constitutive model (equation (4)) is substituted into the balance equation (2) to derive the nonlocal field equation. This field equation is an integropartial differential equation that should be solved for the displacement field, $u_i$.*

For some cases, Eringen's integral constitutive model may give no exact solutions for the nonlocal elasticity problem [19]. As an example, consider a cantilever beam under a point load, $F$, at its free end. The balance equation of the beam's static equilibrium along with the natural boundary conditions can be defined as follows:

$$\frac{d^2 M(x)}{dx^2} = 0 \tag{8}$$

$$\frac{dM(L)}{dx} = -F \text{ and } M(L) = 0 \tag{9}$$

Moreover, according to Eringen's integral constitutive model (equation (4)), the constitutive model of the beam can be obtained in the form:

$$M(x) = D \int_L \alpha(|x' - x|) k(x') \, dx' \tag{10}$$

where $D$ is the constant bending stiffness of the beam, and $k$ is the beam's elastic curvature. $L$ denotes the beam length.

For the considered cantilever beam, Algorithm 1.1 can be used where the nonlocal boundary value problem (equations (8) and (9)) is solved which gives the nonlocal bending moment, $M(x)$, is a linear function:

$$M(x) = F(L - x) \tag{11}$$

To determine the elastic curvature $k$, the obtained nonlocal moment is substituted into equation (10):

$$F(L - x) = \frac{D}{2\ell} \int_L \exp\left(-\frac{|x' - x|}{\ell}\right) k(x') \, dx' \tag{12}$$

It is clear that there is no exact solution for equation (12) since the right-hand side of the equation is nonlinear for any value of the curvature, $k$. This presents a flaw of Eringen's integral constitutive model. However, approximate solutions as the ones proposed in [12-14] are acceptable where the beam curvature can be determined to give a nonlinear moment that can be linearly approximated.

As for nonlocal elasticity problems with essential boundary conditions, Algorithm 1.2 is used where the nonlocal field equation will be obtained in the form of an integropartial differential equation. As stated by Eringen [8], the solution of the integropartial differential equation of the nonlocal elasticity is challenging especially for mixed boundary value problems. This presents another flaw of Eringen's integral constitutive model.

**Remark 1:** *Eringen's integral constitutive model of nonlocal elasticity (equation (4)) gives either a unique solution (which is difficult to be obtained if an integropartial differential nonlocal field equation is formed) or no exact solution (however, approximate solutions are acceptable).*

**Model 2: Nonlocal differential constitutive model**

Because of the complications of solving the integropartial differential nonlocal field equation which is formed when using the integral nonlocal constitutive model (equation (4)), Eringen proposed the differential nonlocal constitutive model [8]. This model is obtained by solving the convolution equation (4) as follows:

$$(1 - \ell^2 \nabla^2) t_{ij} = \sigma_{ij} \tag{13}$$

with the constitutive conditions at the boundary $x = \chi$:

$$\nabla t_{ji}(\chi) + \frac{1}{\ell} t_{ji}(\chi) = 0 \tag{14}$$

To solve the nonlocal field problem (equations (2) and (3)) using Eringen's nonlocal differential constitutive model (equations (13) and (14)), Algorithm 2 is proposed, as follows:

> **Algorithm 2:**
>
> **Algorithm 2.1: Nonlocal field problems with natural boundary conditions**
>
> *The balance equation (2) is solved for the nonlocal stress field, $t_{ij}$, with the nonlocal-type boundary conditions (equation (3)) are applied. Then, the obtained nonlocal stress field is substituted into the differential constitutive model (equation (13)) which is then solved for the local-type stress, $\sigma_{ij}$. The derived solution is acceptable only if it satisfies the constitutive conditions (equation (14)).*
>
> **Algorithm 2.2: Nonlocal field problems with essential boundary conditions**
>
> *The constitutive differential model (equation (13)) is substituted into the balance equation (2). Then, the nonlocal-local boundary conditions relation is formed using the constitutive equation (13) and the constitutive boundary conditions (equation (14)) which can be written in the general form:*
>
> $$[A] \begin{Bmatrix} \sigma_{ij}(\chi) \\ \nabla \sigma_{ij}(\chi) \end{Bmatrix} = \begin{Bmatrix} t_{ij}(\chi) \\ \nabla t_{ij}(\chi) \end{Bmatrix} \quad \text{(a)}$$
>
> *where $[A]$ is the coefficients matrix. Then, the nonlocal field equation is solved for the displacement field, $u_i$. A solution is guaranteed only if Relation (a) is solved.*

A solution of Eringen's differential constitutive model should satisfy the constitutive boundary conditions (equation (14)). Thus, the ability of the differential nonlocal model to give an exact solution for a nonlocal field problem is defined by the fulfillment of the constitutive boundary conditions (equation (14)) [19]. Romano et al. [19] demonstrated that Eringen's differential constitutive model of nonlocal elasticity does not admit a solution for nonlocal elasticity problems because of the violation of this condition. To explain this, the cantilever beam with a point load problem is reconsidered. Thus, the constitutive boundary conditions (equation (14)) will have the form:

$$\frac{dM(0)}{dx} = \frac{1}{\ell} M(0)$$
$$\frac{dM(L)}{dx} = -\frac{1}{\ell} M(L) \tag{15}$$

It is clear that the constitutive boundary condition (15) is violated by the boundary conditions of the cantilever beam (equation (9)). As pointed out by Romano et al. [19] and Mahmoud [24], Eringen's differential constitutive model cannot achieve a consistency between the beam natural boundary conditions (equation (9)) and the constitutive boundary conditions (equation (15)). Therefore, for this case, the differential constitutive model does not admit a solution for nonlocal field problems with natural boundary conditions.

As for nonlocal field problems with essential boundary conditions, Algorithm 2.2 is used in which Relation (a) should be solved. In fact, Relation (a) admits no solution where, in many cases, the coefficient matrix [A] is a singular matrix (*readers can try the case of a cantilever beam with a point load at its free end*).

**Remark 2:** *Eringen's differential constitutive model of nonlocal elasticity (equation (13)) does not admit a solution for nonlocal field problems because either it cannot satisfy the constitutive boundary conditions (equation (14)) or it secretes to a nonlocal-local boundary conditions relation with a singular coefficients matrix. Thus, Eringen's differential constitutive model secretes ill-posed nonlocal boundary value problems.*

**Model 3: Two-phase local/nonlocal constitutive model**

Eringen [2, 21] proposed a nonlocal constitutive model that is a combination of a local-type stress field and a stress field that accounts for the nonlocal effect, as follows:

$$t_{ji}(\boldsymbol{x}) = \xi \sigma_{ij}(\boldsymbol{x}) + (1-\xi) \int_V \alpha(|\boldsymbol{x}' - \boldsymbol{x}|) \sigma_{ij}(\boldsymbol{x}') \, dV(\boldsymbol{x}') \tag{16}$$

This constitutive model is known as the two-phase local/nonlocal constitutive model where it consists of a local phase (the 1st term) and a nonlocal phase (the 2nd term). The weight of each of these two phases can be adjusted via the phase parameter $0 \leq \xi \leq 1$. Thus, when $\xi = 1$, the constitutive model recovers the one of the classical theory. When $\xi = 0$, fully nonlocal constitutive model (equation (4)) is obtained.

The differential form of the integral constitutive model (equation (16)) can be derived as follows:

$$(1 - \ell^2 \boldsymbol{\nabla}^2) t_{ji} = (1 - \ell^2 \xi \boldsymbol{\nabla}^2) \sigma_{ij} \tag{17}$$

with the constitutive boundary conditions at $\boldsymbol{x} = \boldsymbol{\chi}$:

$$\boldsymbol{\nabla} t_{ji}(\boldsymbol{\chi}) + \frac{1}{\ell} t_{ji}(\boldsymbol{\chi}) = \xi \left( \boldsymbol{\nabla} \sigma_{ij}(\boldsymbol{\chi}) + \frac{1}{\ell} \sigma_{ij}(\boldsymbol{\chi}) \right) \tag{18}$$

To solve the nonlocal field problem (equations (2) and (3)) using the two-phase local/nonlocal constitutive models (equation (16) or equation (17)), Algorithm 1 or Algorithm 2 can be used, respectively.

Romano et al. [19, 20] claimed that the ill-posdeness of the fully nonlocal constitutive models (equations (4) and (13)) are eliminated by the two-phase local/nonlocal constitutive model (equation (16) or equation (17)). Thus, the nonlocal elasticity problem is well-posed using the two-phase mixture constitutive model. However, in fact, the two-phase local/nonlocal model admits *either a unique solution or no exact solution the same as the fully nonlocal constitutive model*. Thus, the two-phase constitutive model is also *ill-posed*. To proof this, let us consider the cantilever beam with a point load at its free end. Following Algorithm 1.1, the nonlocal bending model is obtained a linear function as defined in equation (11). According to equation (16), the elastic curvature $k$ can be related to the obtained nonlocal moment as follows:

$$F(L-x) = \xi D k(x) + \frac{D(1-\xi)}{2\ell} \int_L \exp\left(-\frac{|x'-x|}{\ell}\right) k(x') \, dx' \tag{19}$$

It is obvious that equation (19) has no exact solution since the 2nd term of the right-hand side of the equation is nonlinear for any value of the curvature, $k$. This is the same flaw of the fully nonlocal integral constitutive model (equation (4)). Again, approximate solutions, however, are acceptable.

As for the differential form of the two-phase mixture constitutive model (equations (17) and (18)), it can be written for the beam model as follows [19]:

$$M(x) - \ell^2 \frac{d^2 M(x)}{dx^2} = D\left(k(x) - \ell^2 \xi \frac{d^2 k(x)}{x^2}\right) \tag{20}$$

with the constitutive boundary conditions:

$$\frac{dM(0)}{dx} - \frac{1}{\ell} M(0) = \xi D \left(\frac{dk(0)}{dx} - \frac{1}{\ell} k(0)\right)$$

$$\frac{dM(L)}{dx} + \frac{1}{\ell} M(L) = \xi D \left(\frac{dk(L)}{dx} + \frac{1}{\ell} k(L)\right) \tag{21}$$

From the static equilibrium equation (8) of the cantilever beam, the constitutive equation (20) becomes:

$$M(x) = D\left(k(x) - \ell^2 \xi \frac{d^2 k(x)}{x^2}\right) \tag{22}$$

Then, the nonlocal moment and the shear force at the free end of the beam can be determined from equation (22) as follows:

$$M(L) = D\left(k(L) - \ell^2 \xi \frac{d^2 k(L)}{x^2}\right)$$

$$\frac{dM(L)}{dx} = D\left(\frac{dk(L)}{x} - \ell^2 \xi \frac{d^3 k(L)}{x^3}\right) \tag{23}$$

Thus, the left hand side of the second equation of (21) is obtained as follows:

$$\frac{dM(L)}{dx} + \frac{1}{\ell}M(L) = D\left(\frac{dk(L)}{x} - \ell^2\xi\frac{d^3k(L)}{x^3}\right) + \frac{D}{\ell}\left(k(L) - \ell^2\xi\frac{d^2k(L)}{x^2}\right) \qquad (24)$$

It is clear that the right hand sides of equations (21) and (24) are not equivalents. Thus, for the considered cantilever beam, the two-phase local/nonlocal constitutive model (equation (17)) cannot satisfy the constitutive boundary conditions (equation (18)). This conclusion demonstrates that the claims that the nonlocal elasticity problem is well-posed using the two-phase local/nonlocal constitutive model are incorrect.

**Remark 3:** *The nonlocal elasticity problem is ill-posed using the two-phase local/nonlocal constitutive model (equations (16) and (17)). Like the fully nonlocal constitutive model, the two-phase local/nonlocal constitutive model does not admit a solution for nonlocal field problems because it may not satisfy the constitutive boundary conditions (equation (18)).*

## 3. Existence of solutions of nonlocal field problems

Given the foregoing demonstrations and derived remarks, *it can be concluded that the different constitutive models proposed by Eringen secrete nonlocal boundary value problems that admit either a unique solution or no solution at all.* Thus, the paradoxes discussed in [9-18] and the ill-posedness demonstrated in [19, 20] can be attributed to the inconsistency between the nonlocal boundary conditions and the constitutive boundary conditions formed when using any of these constitutive models. Thus, Eringen's constitutive models fail to form the nonlocal stress field and correctly define its values at the boundary of the elastic domain.

Now, the question is: are solutions of nonlocal field problems exist? The author agrees with Romano et al. [19, 20] that an exact solution of an ill-posed nonlocal field problem does not exist at all. Let us accept the fact that any attempt to solve the nonlocal field problem for the nonlocal field using any of Eringen's nonlocal constitutive models is bound to fail (*in exceptional cases, however, Eringen's models may give acceptable solutions*). However, this does not demonstrate that a solution of the nonlocal field problem does not exist! In fact, a nonlocal field problem has a solution! The guarantee to determine this exact solution is constrained by correctly forming the nonlocal field problem. Unfortunately, in many cases, the existing constitutive models cannot correctly form the nonlocal boundary value problem. In this study, a new constitutive model is proposed which guarantees a solution of the nonlocal field problem. Next, this constitutive model is developed.

**Remark 4:** *Solutions of nonlocal field problems are exist. To determine these solutions, a nonlocal constitutive model that can secrete a well-posed nonlocal boundary value problem should be used.*

## 4. Iterative nonlocal residual constitutive model

As previously discussed, the main reason behind the paradoxes and ill-posedness of nonlocal field problems is that Eringen's nonlocal constitutive models cannot correctly form the nonlocal boundary value problem. Therefore, in this section, the iterative nonlocal residual constitutive model is proposed. This new constitutive model can be used to determine a nonlocal solution without solving the original nonlocal boundary value problem. Thus, in the context of this model, the local field equation with an imposed nonlocal residual field is solved for a local-type field. This nonlocal residual is iteratively formed (*not determined*) depending on a pre-determined local field. The iterative nonlocal residual constitutive model permits applying the local boundary conditions. Thus, in the context of this constitutive model, there is no need to form the nonlocal boundary value problem where the local boundary value problem is corrected for the nonlocal field and then solved.

Now, consider an elastic continuum with a volume, $V$, and a surface, $S$, that is exposed to a predetermined residual stress field, $\tau_{ij}$. Hamilton's principle can be then written in the form:

$$\int_0^{t_0} \left( -\int_V \left( \rho \frac{\partial^2 u_i}{\partial t^2} \delta u_i \right) dV + \int_V (\sigma_{ji,j} - \tau_{ji,j}) \delta u_i \, dV - \int_S n_j \sigma_{ji} \delta u_i \, dS + \int_V f_i \delta u_i \, dV \right. \\ \left. + \int_S \bar{P}_i \delta u_i \, dS \right) dt = 0 \tag{25}$$

where $\sigma_{ji}$ is the local stress tensor.

Accordingly, the equation of motion can be written in the form:

$$\sigma_{ji,j} - \tau_{ji,j} + f_i = \rho \frac{\partial^2 u_i}{\partial t^2} \tag{26}$$

and the corresponding natural boundary conditions are:

$$n_j \sigma_{ji} = \bar{P}_i \tag{27}$$

Equations (26) and (27) form *a local boundary value problem* of an elastic domain exposed to a pre-defined residual stress field, $\tau_{ij}$. It should be noted that, because the residual stress $\tau_{ij}$ is known and equation (26) shall be solved for $\sigma_{ij}$, the natural boundary conditions in equations (25) and (27) are the local boundary conditions.

According to the iterative nonlocal residual model, the imposed stress field, $\tau_{ji}$, is a nonlocal-type stress field which is first formed and then imposed to the local field equation (equation (26)). The imposed stress field, $\tau_{ji}$, is iteratively formed depending on a pre-determined local-type stress field. This process permits the application of the local boundary conditions (*no need to form the nonlocal boundary conditions*).

The iterative nonlocal residual constitutive model is defined in equation (28). In the framework of this constitutive model, in an iteration $k$, the nonlocal residual stress, $\tau_{ji}^{(k)}$, is formed based on the pre-determined local stress field of iteration $k - 1$, as follows:

$$\tau_{ji}^{(k)} = \sigma_{ji}^{(k-1)} - t_{ij}^{(k-1)} \tag{28}$$

where $\sigma_{ij}$ is the local-type stress as defined in equation (5). $t_{ij}$ is the nonlocal stress.

Different forms can be proposed for the iterative nonlocal residual constitutive model. Thus, the integral form of this constitutive model can be determined by the substitution of equation (4) into equation (28) which gives:

$$\tau_{ji}^{(k)}(x) = \sigma_{ji}^{(k-1)}(x) - \int_V \alpha(|x' - x|) \sigma_{ji}^{(k-1)}(x') \, dV(x') \tag{29}$$

On the other hand, the differential form of the proposed nonlocal constitutive model can be derived by substituting equation (13) into equation (28), as follows:

$$(1 - \ell^2 \nabla^2) \tau_{ji}^{(k)}(x) = -\ell^2 \nabla^2 \sigma_{ji}^{(k-1)}(x) \tag{30}$$

with the constitutive boundary conditions at $x = \chi$:

$$\nabla \tau_{ji}^{(k)}(\chi) + \frac{1}{\ell} \tau_{ji}^{(k)}(\chi) = \nabla \sigma_{ji}^{(k-1)}(\chi) + \frac{1}{\ell} \sigma_{ji}^{(k-1)}(\chi) \tag{31}$$

To find the solution of a nonlocal field problem using the proposed iterative nonlocal residual constitutive model, Algorithm 3 is proposed, as follows:

| Algorithm 3: |
|---|
| **Algorithm 3.1: Using the Integral Iterative Nonlocal Residual Constitutive Model** |
| 1: **Procedure:** in iteration $k$, |
| 2: (I) Nonlocal residual stress, $\tau_{ji}^{(k)}(x)$, formation: |
| 3: → *Recall $\sigma_{ji}^{(k-1)}(x)$, the determined stress field of the previous iteration $k - 1$.* |
| 4: → *Form $\tau_{ji}^{(k)}(x)$ using equation (29).* |
| 5: → *Save $\tau_{ji}^{(k)}(x)$.* |
| 6: (II) Solving the local boundary value problem: |
| 7: → *Recall $\tau_{ji}^{(k)}(x)$.* |
| 8: → *Form the local boundary value problem:* |
|         $\sigma_{ji,j}^{(k)}(x) - \tau_{ji,j}^{(k)}(x) + f_i = \rho \frac{\partial^2 u_i}{\partial t^2}$    (b) |
|         $n_j \sigma_{ji} = \bar{P}_i$    (c) |
| 9: → *Solve the local boundary value problem ((b) and (c)) for $\sigma_{ji}^{(k)}(x)$.* |
| 10: → *Save $\sigma_{ji}^{(k)}(x)$.* |
| 11: **End Procedure** |

---

**Algorithm 3.2: Using the Differential Iterative Nonlocal Residual Constitutive Model**

1: **Procedure:** in iteration $k$,
2: (I) Nonlocal residual stress, $\tau_{ji}^{(k)}(x)$, formation:
3: $\rightarrow$ Recall $\sigma_{ji}^{(k-1)}(x)$, the determined stress field of the previous iteration $k-1$.
4: $\rightarrow$ Form $\tau_{ji}^{(k)}(x)$ by solving the boundary value problem (equations (30) and (31)).
5: $\rightarrow$ Save $\tau_{ji}^{(k)}(x)$.
6: (II) Solving the local boundary value problem:
7: $\rightarrow$ Recall $\tau_{ji}^{(k)}(x)$.
8: $\rightarrow$ Form the local boundary value problem:
$$\sigma_{ji,j}^{(k)}(x) - \tau_{ji,j}^{(k)}(x) + f_i = \rho \frac{\partial^2 u_i}{\partial t^2} \quad (b)$$
$$n_j \sigma_{ji} = \bar{P}_i \quad (c)$$
9: $\rightarrow$ Solve the local boundary value problem ((b) and (c)) for $\sigma_{ji}^{(k)}(x)$.
10: $\rightarrow$ Save $\sigma_{ji}^{(k)}(x)$.
11: **End Procedure**

---

Algorithm 3 presents a simple procedure to determine the solution of a nonlocal field problem using the proposed iterative nonlocal residual constitutive models. In an iteration $k$, the determined stress field of the previous iteration $k-1$ ($\sigma_{ji}^{(k-1)}(x)$) is used to form the nonlocal residual stress field, $\tau_{ji}^{(k)}(x)$. This nonlocal residual stress is formed by using the integral constitutive model (equation (29)) or by solving the differential constitutive model (equation (30)) with the constitutive boundary conditions (31) are applied. Then, the determined nonlocal residual stress is substituted into the local boundary value problem ((b) and (c)). This local boundary value problem is then solved for the stress field $\sigma_{ji}^{(k)}(x)$ with the local boundary conditions (c) are applied. The process is then repeated where the solution converges to the nonlocal solution with the iterations.

It follows from Algorithm 3 that a solution is guaranteed in each iteration where the local boundary value problem is the one that is solved. The convergence of the obtained solution to the targeted nonlocal solution is constraint by the correct formation of the nonlocal residual stress field. Using the proposed iterative nonlocal residual constitutive models (equations (29) and (30)), the nonlocal residual stress is correctly formed. Moreover, the differential iterative nonlocal residual constitutive model is well-posed where it fulfills the constitutive boundary conditions (equation (31)) for any nonlocal field problem. This is because the differential equation (30) is solved for the nonlocal residual stress with the constitutive boundary conditions (equation (31)) are applied. Thus, the constitutive boundary conditions are automatically fulfilled, and hence the differential constitutive model shall give the same results as the integral constitutive model. This demonstrates that the ill-posedness and the paradoxes associated with Eringen's constitutive model are effectively resolved via the proposed iterative nonlocal residual constitutive model.

**Remark 5:** *Using the iterative nonlocal residual constitutive model, the nonlocal elasticity problem is well-posed and admits a solution.*

## 5. Application to Euler-Bernoulli beams

In this section, a set of nonlocal elasticity problems of Euler-Bernoulli beams are solved using the proposed iterative nonlocal residual constitutive models. Nonlocal elasticity problems of a cantilever beam subjected to a uniform distributed load and/or a point load, a simple supported beam subjected to a uniform distributed load, and a clamped-clamped beam subjected to a uniform distributed load are solved. Solutions of these nonlocal field problems are obtained using the proposed integral iterative nonlocal residual constitutive model and differential iterative nonlocal residual constitutive model.

According to the proposed constitutive model, the elastostatic equilibrium equation of the bending of a beam subjected to a uniform distributed load $q$ can be written as follows:

$$\frac{d^2 M^{(k)}}{dx^2} - \frac{d^2 \mathcal{M}^{(k)}}{dx^2} = q \quad (32)$$

where $M(x)$ is the targeted bending moment. $\mathcal{M}(x)$ is the nonlocal residual bending moment.

The local-type natural boundary conditions can be defined for cantilever, simple supported and clamped-clamped beams, as follows:

**Cantilever beam with a point load at its end:**

$\frac{dM(L)}{dx} = -F$ and $M(L) = 0$

**Simple supported beam:**

$M(0) = 0$ and $M(L) = 0$ \quad (33)

**Clamped-clamped beam:**

$M(0) = \frac{qL^2}{12}$ and $M(L) = \frac{qL^2}{12}$

The nonlocal residual moment, $\mathcal{M}(x)$, can be formed using the integral iterative nonlocal residual constitutive model (equation (29)) as follows:

$$\mathcal{M}^{(k)}(x) = M^{(k-1)}(x) - \int_L \left(\frac{1}{2\ell} \exp\left(-\frac{|x'-x|}{\ell}\right)\right) \left(M^{(k-1)}(x)\right) dx' \quad (34)$$

The corresponding model based on the differential iterative nonlocal residual constitutive model (equation (30)) can be derived as follows:

$$\frac{d^2\mathcal{M}^{(k)}}{dx^2} - \frac{1}{\ell^2}\mathcal{M}^{(k)} - \frac{d^2 M^{(k-1)}}{dx^2} = 0 \tag{35}$$

with the constitutive boundary conditions:

$$\frac{d\mathcal{M}^{(k)}(0)}{dx} - \frac{1}{\ell}\mathcal{M}^{(k)}(0) = \frac{dM^{(k-1)}(0)}{dx} - \frac{1}{\ell}M^{(k-1)}(0)$$

$$\frac{d\mathcal{M}^{(k)}(L)}{dx} + \frac{1}{\ell}\mathcal{M}^{(k)}(L) = \frac{dM^{(k-1)}(L)}{dx} + \frac{1}{\ell}M^{(k-1)}(L) \tag{36}$$

Using Algorithm 3, the targeted nonlocal bending moment, $M(x)$, can be obtained after a few iterations. Afterwards, the deflection of the beam can be determined by solving:

$$\frac{d^2 w(x)}{dx^2} = -\frac{M(x)}{D} \tag{37}$$

with the essential boundary conditions:

**Cantilever beam:**

$w(0) = 0$ and $\frac{dw(0)}{dx} = 0$

**Simple supported beam:**

$w(0) = 0$ and $w(L) = 0$ 

(38)

**Clamped-clamped beam:**

$w(0) = 0$ and $w(L) = 0$

Figure 1 shows the bending moment diagram of cantilever beams subjected to a uniform distributed load and/or a point load, $F$. The results of the proposed constitutive models are obtained for a nonlocal parameter, $\ell = 0.1L$, and compared to the results of the classical beam model. It should be noted that the results of the proposed integral constitutive model (equation (29)) and differential constitutive model (equations (30) and (31)) are identical. The following observations can be derived by inspecting Figure 1:

- *It is clear that the proposed iterative nonlocal residual constitutive models give feasible results where the nonlocal bending field at the free end of the beam vanishes.* As demonstrated by Romano et al. [19], Eringen's constitutive models give infeasible results of cantilever beams where non-zero bending moments were obtained. Thus, the proposed constitutive models can effectively overcome the drawbacks of Eringen's nonlocal constitutive models.

- Using Eringen's constitutive models, the nonlocal bending field was obtained decreasing with an increase in the nonlocal parameters indicating a hardening of the beam stiffness [19]. However, *using the iterative nonlocal residual constitutive models, the nonlocal bending field increases due to nonlocal residuals indicating a softening in the beam stiffness.*

- In various studies [12-14, 19, 20], it was demonstrated that Eringen's differential constitutive model is not an equivalent transformation of Eringen's integral constitutive model. Unlike Eringen's constitutive models, the proposed integral and differential iterative nonlocal residual constitutive models give the exact same results. As previously demonstrated, the differential constitutive model (equations (30) and (31)) is an equivalent transformation of the integral constitutive model (equation (29)). This demonstrate that the proposed constitutive models effective eliminate the paradoxes and discrepancies of nonlocal field problems obtained using Eringen's constitutive models.

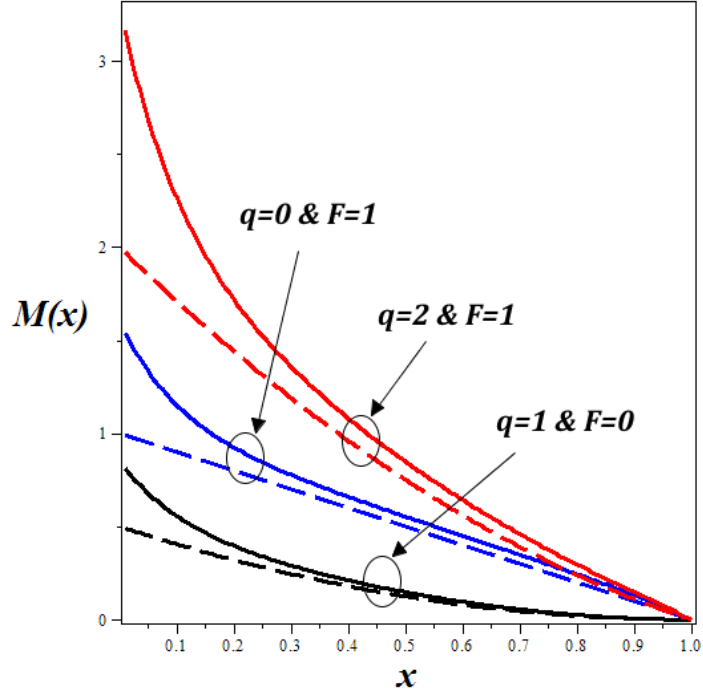

Figure 1: Nonlocal bending moment fields of a cantilever beam ($L = 1$) subjected to a uniform distributed load, $q$, and/or a point load, $F$. Dashed curves denote the local bending fields. The nonlocal bending fields as obtained using the proposed iterative nonlocal residual constitutive models when $\ell = 0.1L$ are presented with solid curves. *The bending moments of the proposed integral and differential constitutive models exactly fit each other.*

Figure 2 shows the nonlocal bending fields of a simple supported beam subjected to a uniform distributed load $q = 1$ as obtained using the integral and differential iterative nonlocal residual constitutive models. Feasible results are obtained as can be observed in Figure 2. The nonlocal bending field at the beam boundaries vanishes. In addition, the nonlocal bending field is obtained with a zero slope at the beam's mid-span. It should be mentioned that the nonlocal bending fields obtained using the integral and differential iterative nonlocal residual constitutive models are identical.

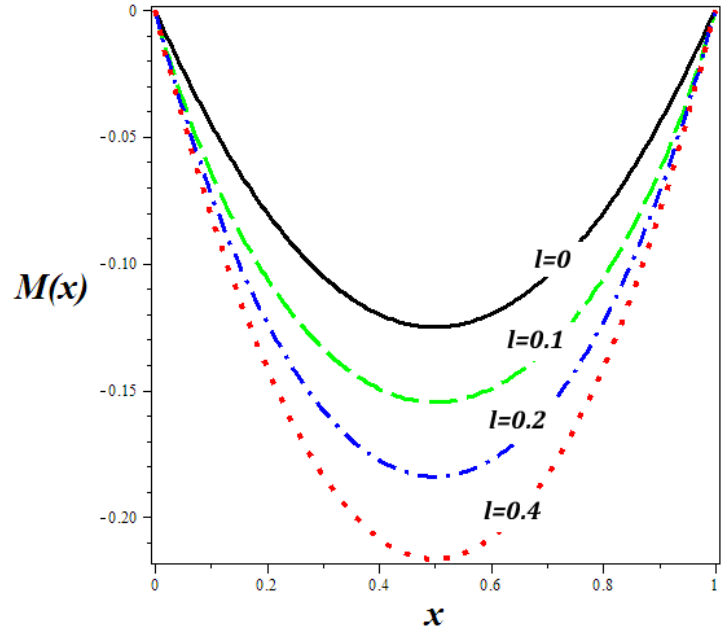

Figure 2: Nonlocal bending fields of a simple-supported beam ($L = 1$) subjected to a uniform distributed load, $q = 1$, for different values of the nonlocal parameter, $\ell = lL$. *The bending moments of the proposed integral and differential constitutive models exactly fit each other.*

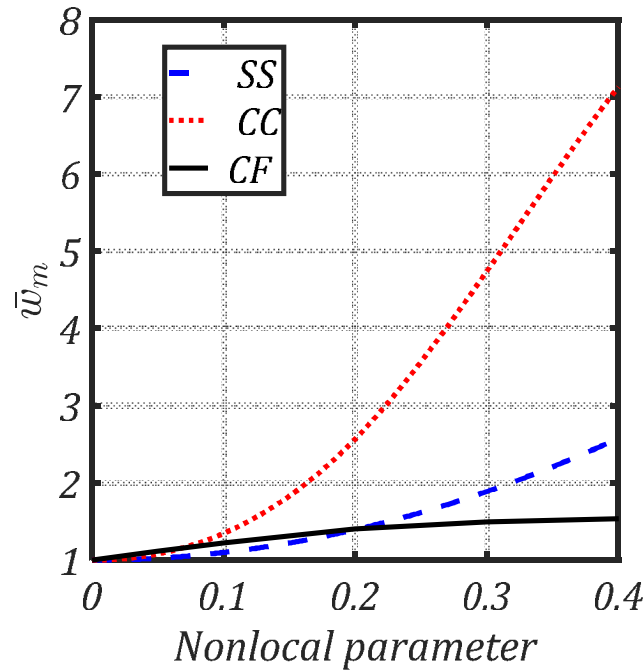

Figure 3: The maximum nonlocal beam deflection-to-the maximum local beam deflection, $\varpi_m$, of an end loaded cantilever beam ($q = 0$, $F = 1$), a simple supported beam under uniform load ($q = 1$), and a clamped-clamped beam under a uniform load ($q = 1$). The results are obtained for different values of the nonlocal parameter, $\ell$, using the proposed constitutive models. *The exact same results are obtained when using the differential and the integral iterative nonlocal residual constitutive models.*

Using the proposed iterative nonlocal residual constitutive models, the maximum nonlocal beam deflection-to-the maximum local beam deflection as a function of the nonlocal parameter for cantilever, simple supported, and clamped-clamped beams is shown in Figure 3. It follows from Figure 3 that the increase in the nonlocal parameter is accompanied with an increase in the beam deflection. Thus, the beam stiffness decreases due to the nonlocal residual fields. It is clear that the paradoxes of cantilever beams are eliminated when using the proposed constitutive models. Again, the proposed integral and differential iterative nonlocal residual constitutive models give the exact same results.

## Conclusions

The integral and differential nonlocal constitutive models proposed by Eringen were discussed and examined according to their ability to secrete nonlocal field problems that admit solutions. Eringen's integral and differential constitutive models of fully nonlocal and two-phase local/nonlocal elasticity were examined. The performed analyses revealed:

- All Eringen's nonlocal constitutive models likely give nonlocal field problems that either are unsolvable or admit no exact solutions.
- Eringen's integral nonlocal constitutive models (equation (4) and equation (16)) secrete integropartial differential nonlocal field equations that do not admit exact solutions. However, approximate solutions may be obtained.
- Eringen's differential constitutive model of the fully nonlocal elasticity (equation (13)) does not admit a solution for nonlocal elasticity problems because either it cannot satisfy the constitutive boundary conditions (equation (14)) or it secretes a nonlocal-local boundary conditions relation with a singular coefficients matrix.
- The two-phase local/nonlocal constitutive model (equations (16) and (17)) secretes ill-posed nonlocal elasticity problems. On the contrary to Romano's [19, 20] claims of its well-posedness, it was demonstrated that the two-phase local/nonlocal constitutive model, like the fully nonlocal constitutive model, does not admit a solution for nonlocal field problems because it may not fulfill the constitutive boundary conditions (equation (18)).
- Solutions of nonlocal field problems are exist. To determine these solutions, a nonlocal constitutive model that can secrete a well-posed nonlocal boundary value problem should be used.

To remedy the drawbacks of Eringen's constitutive models and to correctly form the nonlocal boundary value problem, the iterative nonlocal residual constitutive model was developed. In the context of this model, the local boundary value problem with an imposed nonlocal residual field is solved where the local-type boundary conditions are applied. An algorithm that explains an iterative procedure to form the nonlocal residual and to iteratively correct the local boundary value problem for the nonlocal residual was proposed. The integral and differential forms of the proposed constitutive model were derived (equations (29) and (30)). The proposed constitutive models were examined according to their ability to secrete well-posed nonlocal boundary value problems. Case studies of nonlocal elasticity problems of Euler-Bernoulli beams with differential natural and essential boundary conditions were carried out. The results came to demonstrate the following conclusions:

- The iterative nonlocal residual constitutive model secretes well-posed nonlocal boundary problems.
- A solution of a nonlocal elasticity problem is guaranteed when using the iterative nonlocal residual constitutive model.
- The paradoxes and ill-posedness of Eringen's constitutive models are effectively eliminated when using the iterative nonlocal residual constitutive model.
- The constitutive boundary conditions of the iterative nonlocal constitutive model are automatically fulfilled.
- The differential form of the iterative nonlocal residual constitutive model is equivalent to its integral form where both models secrete the exact same solutions of nonlocal field problems.

In conclusion, any attempt to solve nonlocal elasticity problems using any of Eringen's constitutive models is bound to fail where ill-posed nonlocal boundary value problems are secreted. In contrast, a solution of nonlocal elasticity problems is guaranteed using the iterative nonlocal residual constitutive model which secretes well-posed nonlocal boundary value problems.